\title{Arbitrage Opportunities in CDS Term Structure: Theory and  Implications for OTC Derivatives} 
\author{Raymond Brummelhuis\thanks{Dept. of Mathematics \& Computer Science, Univ. of Reims, France, raymondus.brummelhuis@univ-reims.fr}
        \and
        Zhongmin Luo\thanks{Dept. of Economics, Mathematics \& Statistics, Birkbeck, Univ. of London, UK, 
         zhongmin.luo@btinternet.com.}
        }
\documentclass[12pt]{article}
\usepackage{graphicx}
\usepackage{latexsym,amsmath}
\usepackage{amsfonts} 
\usepackage{amssymb}
\usepackage{amsthm}
\usepackage{pifont} %% for \ding
\usepackage{hhline}
\usepackage{dsfont}
\usepackage{epsfig}
\usepackage{rotating}
\usepackage{url}
\usepackage{caption}
\usepackage{makeidx}
\usepackage{bm}
\usepackage{hyperref}
\makeindex
\usepackage{txfonts} 
\usepackage{pxfonts}
\usepackage[titletoc,title]{appendix}
\usepackage{mathtools}

\usepackage[top=1in, bottom=1in, left=1in, right=1in]{geometry}
\usepackage{geometry}   
\usepackage{color}  
\usepackage{xcolor}

\theoremstyle{definition}
\newtheorem{remark}{Remark}

\vbadness=\maxdimen

\newtheorem{theorem}{Theorem}

\newtheorem{proposition}{Proposition}   
\newtheorem{corollary}{Corollary}   
\newtheorem{examples}{Examples}   
\newtheorem{definition}{Definition}

\def\bleu{}
\def\noir{}   

 \usepackage{setspace,caption}  %set up double space
\usepackage{color}   
\usepackage{setspace,caption}   
  
\begin{document}
\newpage
\maketitle
\begin{abstract} 
%\marginpar{\tiny\textcolor{red}{12Nov: I have shortened the Abstract as suggested.}} 
 
Absence-of-Arbitrage (AoA) is the basic assumption underpinning derivatives pricing theory. As part of the OTC derivatives market, the CDS market not only provides a vehicle for participants to hedge and speculate on the default risks of corporate and sovereign entities, it also reveals important market-implied default-risk information concerning the counterparties with which financial institutions trade, and for which these financial institutions have to calculate various valuation adjustments (collectively referred to as XVA) as part of their pricing and risk management of OTC derivatives, to account for counterparty default risks. In this study, we derive No-arbitrage conditions for CDS term structures, first in a positive interest rate environment and then in an arbitrary one. %, and show that violating such conditions leads to negative or greater-than-1 conditional default probabilities and CDS-term-structure arbitrage opportunities.   
Using an extensive CDS dataset which covers the 2007-09 financial crisis, we present a catalogue of 2,416 pairs of anomalous CDS contracts which violate the above conditions. Finally, we show in an example that such anomalies in the CDS term structure can lead to persistent arbitrage profits and to nonsensical default probabilities. The paper is a first systematic study on CDS-term-structure arbitrage providing model-free AoA conditions supported by ample empirical evidence.    
%it attributes the existence of arbitrage opportunities to implicit costs and risks borne by would-be arbitrageurs such as counterparty risks, liquidity risks, etc., for which an array of valuation adjustments have been developed as part of OTC derivatives pricing and risk management. The study argues that successful efforts in valuation adjustments amongst others may reduce of the amount of arbitrage opportunities in the CDS market, which is conducive to the health of OTC derivatives market. 
%The paper concludes with future research directions indicated.   
\medskip   
   
 \end{abstract}

\section{Introduction}\label{sectionIntroduction}   
   
Derivative pricing theory has as its basic assumption that the financial markets are arbitrage-free, in the sense that there should not exist an opportunity to make a risk-free profit without initial setup costs.  With an appropriate mathematical definition (a non-trivial matter, especially in a continuous-time framework: see for example Delbaen and Schachermayer (2006, \cite{Schachermayer}), Absence-of-Arbitrage can be shown to be equivalent to the existence of an equivalent martingale measure with respect to which prices can be evaluated as expectations of discounted pay-offs: see for example Bjork \cite{Bjork} (2005).    

The Over-The-Counter or OTC derivatives represent a market with a notional of US\$532 trillion (BIS Statistics \cite{BISstats}, 2017), including derivatives on foreign exchange, interest rates, commodities, credit, equities and other derivatives. As part of OTC derivatives market, the Credit Default Swap (or CDS) not only allows participants to speculate on and hedge credit risks, it also reveals market-implied default-risk information about financial institutions' counterparties. The latter is critical for the pricing and risk-management of OTC derivatives (especially for non-centrally cleared ones), in order to properly account for counterparty credit risk in the form of Credit Valuation Adjustment or CVA and other valuation adjustments collectively referred to as XVA: see for example Brigo, Morini and Pallavicini \cite{Brigo} (2013). One of the main lessons learned from the 2007-09 financial crisis is that the market failed to properly account for counterparty default risks in the pricing and risk management of OTC derivatives. A lot of progresses have been made in this area since the crisis\footnote{See for example Brigo, Francischello and Pallavicini for non-linear valuation adjustment pricing framework (2015, \cite{BrigoNVA}), Crepey \textit{et al} (2013, \cite{Crepey}) for a BSDE-based (Backward stochastic differential equation) approach and Burgard and Kjaer (2013,\cite{Burgard}) for a replicating approach for FVA; Green, Kenyon and Dennis (2014, \cite{GreenKenyonDennis}) and Green and Kenyon (2015, \cite{GreenKenyon}) among others extended the replicating approach to include capital and margin requirements in forms of KVA (Capital Valuation Adjustment) and MVA (Margin Valuation Adjustment) respectively.}. In practice, the post-crisis pricing and risk management of OTC derivatives in banks consist of two parts, risk-neutral valuation of the derivative by the  Front Office assuming there is no counterparty-default- risk,  and the addition of an array of valuation adjustments conducted by the XVA Desk, also computed by risk-neutral valuation in risk-neutral world. Valuation adjustments have become standard practice in pricing and risk management of OTC derivatives for most large banks. 
 
\smallskip

Literature often highlights that a CVA can be seen as a Contingent CDS or CCDS (cf. Brigo, Morini and Pallavicini \cite{Brigo}, 2013), which pays the Loss-Given-Default on the residual net present value of a given portfolio. A CCDS is a type of credit derivatives whose notional and maturity are contingent on the time-of-default of an underlying reference entity. To price the CVA or CCDS, the default risk of the reference entity has to be calibrated to liquid CDS rates in a presumably arbitrage-free market. However, this element of the CVA pricing has not received much attention; essentially, the literature typically assumes that the counterparties of financial institutions either have liquidly quoted CDSs contracts on their names, or that sound proxies are available and that the CDS market is arbitrage-free. However, as regards the former assumption, as highlighted in Brummelhuis and Luo (2018a) \cite{2018a},  the vast majority of counterparties of financial institutions are not liquidly quoted in the CDS market and CDS proxy rates  have to be constructed. In that reference, a CDS-proxy construction method via Machine Learning techniques has been proposed and shown to be more sound and to produce more accurate results than existing proxy methods: see Brummelhuis and Luo (2018b) \cite{2018b}, for a benchmarking exercise. Regarding the latter, as shown below, the CDS market is not as arbitrage-free as assumed in derivatives pricing theory; in fact, a significant amount of CDS-term-structure arbitrage opportunities exist in the CDS market. Alternatively, interested readers can find a shorter version of the paper for ML-based CDS Proxy Construction in Brummelhuis and Luo (2017 [12]). 
\medskip   
   
Using simple arguments we show that, for CDS contracts of varying maturity on a given name,  the CDS-spread times the maturity has to be an increasing function of the maturity. This can be shown without making any modeling assumptions if the risk-free interest rates are positive. When risk-free rates can be negative we derive a version of this theorem in which we multiply the CDS-spreads not by the maturity, but by what we will call the standardized risk-free annuity, which is the present value of an annuity paying a standardized interest rate of 1 over the life-time of the CDS contract (at the same payment dates as the CDS). In the case of positive interest rates this strengthens the first condition above, but its derivation assumes the existence of an equivalent martingale measure or risk-neutral pricing measure\footnote{For finite period markets the existence of such a measure is equivalent to absence of arbitrage, but for general continuous time markets in which prices are semi-martingales the latter has to be strengthened to for example the NFLVR-condition of Delbaen and Schachermayer (1994) \cite{Delbaen}; in this paper we will generally disregard such subtleties.} and the conditional independence of default-times of the other market variables. These assumptions also underlie current standard practice in banks for deriving market-implied (or risk-neutral) default probabilities from CDS quotes by the so-called Bootstrapping approach (see for example O'Kane (2008), \cite{Okane1}) as well as the evaluation of CVAs. We then found that these absence-of-arbitrage conditions can be violated in CDS markets: see subsection \ref{sectionAnomalies} and section \ref{sectionEmpirics} below.     
\bigskip   
   
\smallskip

Despite the uniquely important role played by the CDS market, there are only a limited number of studies on CDS-related arbitrage: see Kapadia and Pu (2012, \cite{Kapadia}) for an example regarding arbitrage between CDS and equity markets and Bai and Collin-Dufresne (2012, \cite{Baietal}) for an example on  arbitrage between CDS and bond markets. A recent study by Jarrow \textit{et al} (2018, \cite{Jarrow2}) focused on statistical arbitrages in the CDS term structure by identifying mis-pricing of CDSs based on the gaps between theoretical CDS prices and observed ones; their results however are model-dependent, contrary to ours.

\smallskip
   
There are potential implications of CDS-term-structure arbitrage for the pricing and risk management of both credit derivatives and general OTC derivatives, because of the requirement to compute  the valuation adjustments. Given the importance of the No-arbitrage assumption in deriving default probabilities from observed CDS curves in a presumably efficient CDS market, it is important to understand to what extent this market is arbitrage-free. Do we see potential arbitrage opportunities and are these being taken advantage of by arbitrageurs? As we will see, the answers to this are, respectively, a "yes" and a "no". These questions are of fundamental importance for the pricing and risk management of credit derivatives and XVA.
   
In the remainder of this introductory section we first present an overview of anomalous CDS trades which violate the simplest of our No-arbitrage conditions and which motivated our study. We then more formally list the research gaps we identified and the contributions we believe this paper makes to the existing literature.   
       
\subsection{Anomalous CDS term structures}\label{sectionAnomalies}
    
Let  $s_1 $ and $s_2 $ be the mid-point CDS spreads of two co-initial standard single-name CDS contracts starting at $T_0 $ with maturities $T_1 < T_2 $ but otherwise being identical\footnote{It means identical vintage, underlying bond, seniority type, currency and notional, etc. in ISDA master agreement.}. According to Theorem \ref{thm:NAC_1} in section \ref{sectionNoarbitragecondition}, Absence-of-Arbitrage (AoA) in the CDS market implies that: 
     
\begin{equation}\label{eqNoarbitrageCondition}   
(T_1-T_0)s_1%(T_0, T_1) 
< (T_2-T_0)s_2 . %(T_0,T_2).   
\end{equation}   
We will call the pair of CDS trades for which (\ref{eqNoarbitrageCondition}) is violated an {\it Arbitrage Trade} or more briefly an {\it Anomaly}. Note that in that case, necessarily $s_1 > s_2 $, so the CDS curve is inverted. Inverted yield curves are well-studied by researchers but inverted CDS curves have not attracted much attention from researchers. O'Kane (2008, \cite{Okane1}) remarked that model-independent arbitrage may exist in the CDS market and that negative implied default probabilities can occur for inverted CDS curves. He stated a condition similar to (\ref{eqNoarbitrageCondition}) but in a form and with a derivation which suggests it is only holds approximatively, while it is in fact exact. Brigo and Mercurio \cite{BrigoIR} (2006) also commented on the possibility of negative hazard rates as a result of inverted CDS curves. However, no systematic study has been conducted to assess the significance of this an issue and its implications for OTC derivatives.

\smallskip

In the case of an anomaly violating (\ref{eqNoarbitrageCondition}) as described above, an arbitrageur would simultaneously sell the protection against the default of an entity ("Short  the CDS") for a rate $s_1 $ over the period of $T_1-T_0$ and buy the default protection against the same entity ("Long the CDS") for a rate of $s_2 $ over the period $T_2-T_0$. We 
refer to this as \textit{a paired arbitrage trade}, denoted informally by $\left( (s_1 , T_1 ) ; (s_2 , T_2 ) \right) . $   
\medskip

We now turn to identify potential arbitrage opportunities by checking a list of paired trades $\left( (s_1 , T_1 ) ; (s_2 ; T_2 ) \right) $ against condition (\ref{eqNoarbitrageCondition}) with the focus on single-name CDS contracts traded between January of 2007 and June of 2010, which includes the official beginning and end of the 2007-09 crisis   
\footnote{See National Bureau of Economic Research: \url{http://www.nber.org/cycles.html} for a definition of US economic cycles}. For this study, we have sourced CDS data from Bloomberg\texttrademark and Thomson Reuters.  

Our data sample is comprehensive in that it contains over 3 million CDS curves, each with 8 CDS contracts of different maturities (i.e., 6-month, 1-year, 2-year, 3-year, 4-year, 5-year, 7-year and 10-year) traded over 868 consecutive trading dates\footnote{We avoid  data manipulations such as term interpolations to ensure that our data truly reflect the CDS market.}. From these CDS trade data, we then identify a list of \textit{paired CDS trades} that violate condition (\ref{eqNoarbitrageCondition}) with focus on the following four paired maturities: $(T_1, T_2 ):= \{(0.5, 1 ) $, $(1, 2 ) $, $(2, 5 ) $, $(5,10)\}$, where maturities are measured in years. We therefore note that our analysis in terms of paired maturities is not exhaustive. 
\medskip

Table \ref{figureSummaryofCountryRatingProfile} presents a summary of the number of anomalies (indicated below in brackets) based on the following five attributes: (\textit{i}) across three major \textit{regions}, i.e., Asia (42), Europe (295) and North America (2,079); (\textit{ii}) across \textit{sector groups}, i.e., Banking (187) and Non-banking (2,229); (\textit{iii}) across three major \textit{currencies}, i.e., USD (1,836) and EUR (579) and GBP(1); (\textit{iv}) across \textit{rating class groups}, i.e., Investment-grade (1,863), Non-investment grade or NIG (180) and Not Rated (373); (\textit{v}) across \textit{seniority types} for the debt obligation underlying a CDS trade, i.e., Senior (2,385), Senior Secured (12) and Subordinated (19). We note that   

\begin{itemize}

\item As regards regions, the North American region has higher percentage (86\%) of anomalies than others. We believe this is mainly due to two reasons: in our CDS data sample, 64\% of CDS trades have their underlying entities operating in this region; furthermore, the sample  period of our CDS data includes the on-going financial crisis that affected the North American region more than the others regions, which may have contributed to the observed higher proportion of anomalies for the region.  

\item Regarding sector groups, the 2,229 anomalies in the Non-banking sector corresponds to 92.3\% of the total number of anomalies.

\item As for rating groups, 1,863 or 77.1\% anomalies are associated with Investment-grade debt, leaving the remaining non-negligible number of anomalies associated with either Non-investment-grade or Not-rated debt. For reasons which are further discussed in section \ref{sectionEmpirics}, non-negligible numbers of anomalies  are associated with either AAA-rated (36) or AA-rated (222) names.  

\item With regard to seniority, almost all anomalies are associated with either \textit{senior} or \textit{senior secured} debt, suggesting debt seniority type (junior versus senior) may not be a determining factor explaining the existence of anomalies in our sample. 

\end{itemize} 

\begin{table}
%\vspace{-5mm}
\caption{\# of Anomalies by Regions, Sectors, Currencies, Credit Ratings and Seniority Types}\label{figureSummaryofCountryRatingProfile} 
\centering  
\includegraphics[scale=.75]{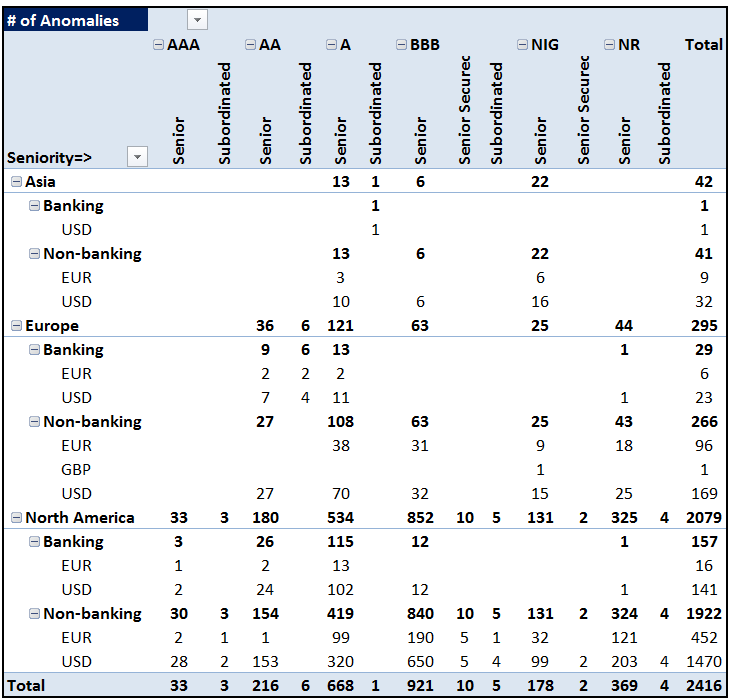} 
%\vspace{-1mm}
\end{table}

To sum up, we have presented a cross-sectional overview of CDS-term-structure arbitrage opportunities identified in this study; in section \ref{sectionEmpirics}, we present further evidences of such anomalies across time and discuss the implications.
    
\subsection{Research Gaps and Contributions} \label{sectionLiterature}      

Before we move to the next section, we present a summary of the research gaps which motivated our study and list the contributions we believe this study makes to literature.   
 
\subsubsection*{Research gaps}
We identified the following research gaps in existing literature:

\begin{enumerate}
\item The existing literature valuation adjustment literature assumes that the CDS market, which is the basis for the calibration of counterparty default probabilities, is arbitrage-free. Given the importance of valuation adjustment calculation for OTC derivatives, it is important to investigate whether the CDS market is truly arbitrage-free; if not, it is important to understand how significant the arbitrage opportunities are, how persistent such opportunities are, why would-be arbitrageurs are not taking advantage of them and what the implications are for OTC derivatives pricing and risk management.  

\item Past CDS-arbitrage related literature focused on either CDS-bond arbitrage or CDS-equity arbitrage. No systematic study has been conducted to identify and explain CDS-term-structure arbitrage except for a recent paper by Jarrow \textit{et al} (2018, \cite{Jarrow2}), where, however, the authors focused on statistical arbitrage using a model-dependent approach.   

\item Although O'Kane \cite{Okane1} (2008) and Brigo and Mercurio \cite{BrigoIR} (2006) commented on the potential occurrence of negative hazard rates in the calibration of default probabilities to CDS curves, no systematic study of arbitrage opportunities in the CDS term structure and of its implications for the pricing and risk management of OTC derivatives has been conducted to date. For example, there does not seem to exist a basically model-free mathematical criterion for a CDS curve to be arbitrage free.  
\end{enumerate}

\subsubsection*{Contributions}
We summarize the main contributions of our paper:   
   
\begin{enumerate}
\item The paper represents the first systematic study of CDS-term-structure arbitrage on the basis of a simple model-independent criterion for No-arbitrage, supported by empirical evidence and numerical examples based on real data.   
   
\item The No-arbitrage conditions we derive are basically model-independent, assuming at most the existence of an equivalent martingale measure, and sometimes not even that.     

\item It shows that violating these conditions can lead to either negative or greater-than-1 conditional default probabilities, which has implications for the pricing and risk management of both credit derivatives and OTC derivatives in general, through the computation of value adjustments such as CVA and DVA.   

\item It argues that an explanation of these apparent arbitrage opportunities can be found in the presence of implicit risks and costs which were systematically neglected in the practice of OTC-derivatives pricing prior to the crisis of 2008-9 and which have since become the focus of research efforts of academics,  and practitioners through the inclusion of various valuation adjustments, and that taking these into account in the calculation of CDS spreads should reduce arbitrage opportunities in the CDS market. 
\end{enumerate}
 
The rest of the paper is organized as follows: in section \ref{sectionNoarbitragecondition}, we derive the No-arbitrage condition (\ref{eqNoarbitrageCondition}) for a CDS term structure with zero modelling assumptions, under the assumption that risk-free rates are non-negative. We do this first (by way of example) for a simple three period model, and then for the continuous time case with both continuous and discrete protection payments. 

In section \ref{sectionAllRates}, we use the existence of an Equivalent martingale Measure to derive a No-arbitrage condition for a general interest rate environment, with potentially negative short rates, which generalizes (\ref{eqNoarbitrageCondition}) if interest rates are positive. In section \ref{sectionEmpirics} we present further empirical evidence for CDS-term-structure anomalies and the existence of arbitrage profits. We present numerical examples of the implications of such anomalies for conditional default probabilities. Finally, section \ref{sectionConclusion} concludes.

\section{\bf No-arbitrage conditions on CDS Term Structure }\label{sectionNoarbitragecondition}

We consider a co-initial family of CDS contracts providing protection against the default of a reference entity between times $T_0 $ and $T $ for arbitrary $T > T_0 . $ The default protection \textit{seller} pays the \textit{buyer} the Loss-Given-Default should the default occur within $[T_0, T ] $ in exchange for either a continuously or discretely paid spread of  $s_t (T_0 ; T )$, which is fixed at the inception of the contract and paid by the \textit{buyer} up till time-of-default $\tau$ or up till the contract's maturity $T $, whichever is smaller. Here $t < T_0 $ is the time at which the spread is agreed upon between the two parties of the CDS. We are interested in potential static arbitrage opportunities presented by the CDS term structure or CDS curve $\{ s_t (T_0 , T ) : T \geq T_0 \} $ at any fixed time $t $,  which, without loss of generality, we can take to be $t = 0 $ (corresponding to "today" or "now"). Dynamic No-arbitrage conditions for the stochastic evolution of these curves as function of $t $ are outside of the scope of the present paper.   
\medskip   
   
\subsection{No-arbitrage condition for discretely paid CDS rates: a simple example}      
We first look at a simple discrete time model, with times $0, 1, 2 $ and $3 . $ To further simplify, we assume that interest rates are 0 (this will be relaxed in the next section). Suppose that the defaultable bond underlying the CDS contract is a 0-coupon bond with maturity $T = 3 $ and nominal value 1, and that the reference entity has a strictly positive risk-neutral probability of default before maturity. We assume of course that the entity has not yet defaulted at the trade date $t = 0 $ of  the contract: $\mathbb{P } (\tau > 0 ) = 1 . $ Consider two CDS contracts, $CDS_1 $ and $CDS_2 $, both starting at time $t = 1 $ (the Effective Date of the CDS) 
with $CDS_1 $ providing protection against default at $t = 2 $ for a spread of $s_1 $, and the second contract, $CDS_2 $, providing protection against default at times $t = 2 $ and $t = 3 $ for a spread of $s_2. $ The spreads are agreed upon at time 0, and therefore known. Let $R_{\tau } \in [0 , 1 ) $ be the, possibly random, recovery-upon-default at the time-of-default, which we assume to be strictly positive.     
 
Suppose that $s_1 \geq s_2 $, corresponding to an inverted CDS curve with short-term protection being more expensive than long-term one, and that we short the first CDS, in the sense  that we sell protection against default at time $T_2 $ for a fee of $s_1 $ payable\footnote{We follow the convention that protection payments for a given period are done at the end of the period.} at $T_2 $, and go long the second one, so that we are buying default protection against a fee of $s_2 $ payable at $T_2 $ and $T_3 $, assuming no default has yet occurred at the payment times. Our initial investment at time 0 (initial net cashflow) is 0, since the contracts have not started yet. If $\tau = 1 $, both contracts are cancelled on their Effective Date. If $\tau > 1 $, we receive $s_1 - s_2 $ at time 2. If default happens at time 2, the payments-upon-default of the two CDS contracts cancel out, while if $\tau > 2 $ we have to pay $s_2 $ at $T = 3 . $ If $\tau = 3 $, we receive $1 - R_3 $ from the protection seller of the second contract.   These are all the possibilities for $\tau $. Assuming we carry along all received payments in a savings account, our net accumulated cash-flow at time 3 then equals
$$
(s_1 - s_2 ) \mathbf{1 }_{\{ \tau > 1 \} } - s_2 \mathbf{1 }_{\{ \tau > 2 \} } + (1 - R_3  ) \mathbf{1 }_{\{ \tau = 3 \} } \geq (s_1- 2 s_2 ) \mathbf{1 }_{\{ \tau > 1 \} } + (1 - R_3 )  \mathbf{1 }_{\{ \tau = 3 \} } .
$$
Hence we have an arbitrage opportunity if   
\begin{equation} \label{eq:discrete_model_AO}
s_1- 2 s_2 \geq 0 ,    
\end{equation}   
assuming that $\mathbb{E } ((1 - R_3 ) \mathbf{1 }_{\tau = 3 } ) > 0 $ (if not, then there will always be an arbitrage if $s_1 > 2s_2 $).    
Note that, in fact, the net value of the position is non-negative at all intermediary times.    

\begin{remark}\label{remark1} \rm{If the CDS spread curve is not inverted, $s_1 \leq s_2 $, then by shorting the CDS with the higher spread and going long the other CDS, the net amount we will have received at time 3 is $(s_2 - s_1 ) \mathbf{1 }_{\{ \tau > 1 \} } + s_2 \mathbf{1 }_{\{ \tau > 2 \} } - (1 - R_3 ) \mathbf{1 }_{\{ \tau = 3 \} } $: 
the situation is different from the previous case of inverted spreads, since $1 - R_3 $ now occurs with a minus sign.   
\smallskip

}
\end{remark}

It is instructive to examine the above example from the viewpoint of risk-neutral pricing. Let $\mathbb{Q } $ be an equivalent martingale measure and suppose that the conditional risk-neutral probabilities of default over  each period are constant:     
$$
\mathbb{Q } (\tau = i \vert \tau > i  - 1 ) = \lambda, \ \ i = 1, 2, 3 ,   
$$   
while $\mathbb{Q } (\tau > 0 ) = 1 . $   
Furthermore, assume that the loss-given-defaults $L_t > 0 $ at $t = 2, 3 $ are constant; if the reference entity for the CDS is a zero-coupon bond of face value 1, then $L_t = 1 - R_t . $ We will see that if (\ref{eq:discrete_model_AO}) holds, then 
$\lambda $ either has to be negative or strictly bigger than 1. The fair value at time 0 of the first spread is determined by   
$$
s_1 \mathbb{Q } (\tau \geq 2 ) = L_2 \mathbb{Q } (\tau = 2 ) \Leftrightarrow s_1 = L_2 \lambda ,   
$$
since $\mathbb{Q } (\tau \geq  2 ) = \mathbb{Q } (\tau \geq 2 | \tau > 1 ) \mathbb{Q } (\tau > 1 | \tau > 0 ) = 1 \cdot (1 - \lambda ) = (1 - \lambda ) $ and, similarly, $\mathbb{Q } (\tau = 2 ) = \lambda (1 - \lambda ) . $    

The equation for the fair value of the second spread is
$$
s_2 \left( \mathbb{Q }(\tau \geq 2 )+ \mathbb{Q } (\tau \geq 3 ) \right) = L_2 \mathbb{Q } (\tau = 2 )+ L_3 \mathbb{Q } (\tau = 3 ),    
$$
or $s_2 \left( (1 - \lambda ) + (1 - \lambda )^2 \right) = \lambda (1 - \lambda ) L_2 + \lambda (1 - \lambda )^2 L_3 $, so that   
$$
s_2 = \lambda \frac{L_2 + (1 - \lambda ) L_3 }{2 -\lambda } .
$$
Observe that if $L_1 = L_2 = L $, then $s_1 = s_2 = \lambda L $, and 
condition (\ref{eq:discrete_model_AO}) would imply that $s_1 - 2 s_2 = - \lambda L<0 $, or $\lambda < 0 $, and our risk-neutral pricing model for pricing the CDS would break down.   

More generally, letting $x := L_3 / L_2 $, we find that   
$$
s_1- 2 s_2 = - \lambda L_2 \frac{\lambda + 2 (1 - \lambda ) x }{2 - \lambda } ,   
$$
and the condition that $s_1- 2s_2 > 0 $ forces $\lambda $ to be either strictly smaller than 0 or bigger than 1. (If $\lambda > 0 $, then the second factor of the product on the right has to be strictly negative, which forces $\lambda > 1 $, while if $\lambda < 1 $, $\lambda $ has to be negative because of the first factor; note that $x > 0 $ since $L_1 $ and $L_2 $ are.)   
\medskip

Another way of deriving that Absence of Arbitrage implies that $s_1 \leq 2 s_2 $ is by observing that the risk-neutral expectation at time 0 of the pay-off of the difference of the two CDS contracts ("short $s_1 $, long $s_2 $") is 
%\marginpar{\small\textcolor{violet}{ZL-26August: similarly, I added $1-R_3$ in the formula here}}
$$
(s_1 - s_2 ) \mathbb{Q } (\tau > 1 ) - s_2 \mathbb{Q } (\tau > 2 ) + L_3 \mathbb{Q } (\tau = 3 ) .
$$
Since this has to be 0, and since the last term is positive, it follows that
$$
(s_1 - s_2 ) \mathbb{Q } (\tau > 1 ) - s_2 \mathbb{Q } (\tau > 2 ) \leq 0 ,   
$$
or
\begin{equation} \label{eq:discrete_AoA}
s_1 \leq \left( 1 + \frac{\mathbb{Q }(\tau > 2 ) }{\mathbb{Q } (\tau > 1 ) } \right) s_2 = \left( 1 + \mathbb{Q } (\tau > 2 | \tau > 1 ) \right) s_2 ,   
\end{equation}
which certainly implies that $s_1 \leq 2 s_2 $, since $\mathbb{Q }(\tau > 2 ) \leq \mathbb{Q } (\tau > 1 ) . $ Note that (\ref{eq:discrete_AoA}) is sharper than $s_1 \leq 2s_2 $ but that the latter makes no reference to survival probabilities. Inequality (\ref{eq:discrete_AoA}) would be useful when we would have access to survival probabilities from another source than the CDSs themselves, e.g. from defaultable bond yields. 

\subsection{\bf No-arbitrage conditions for continuously paid CDS rates}

We now put ourselves in a continuous-time framework, and consider a family of co-initial forward CDS contracts providing default-protection over a future time periods $[T_0 , T ] $, with $T_0 \geq 0 $ fixed and $T > T_0 $ arbitrary. We let $\tau $ be the time of default.   
The party which is long the CDS (the buyer of the default protection) will pay a protection fee between $T_0 $ and $T $ up till default of the reference entity, in which case he will receive the Loss-given-default (LGD). We will consider both the case of a protection fee which is paid out continuously and one which is paid discretely, at payment dates 
$$   
t_0 = T_0 < t_1 < \cdots < t_{n - 1 } < T .   
$$   
In the first case the protection buyer pays a continuous fee at a rate of $s dt $ up till time-of-default, and in the second case he pays $s (t_i - t_{i - 1 } ) $ at time $t_i $ for $i = 1 , \ldots , n $, with $t_n := T $ if $t_i \leq  \tau $, while he only pays the spread over $[ t_{i - 1 } , \tau ] $ if default $\tau $ occurs between $t_{i - 1 } $ (excluded) and $t_i $ (included). Here the {\it CDS spread} $s = s_0 (T_0, T ) $ is determined at time 0 so that if $T_0 > 0 $ it is in fact a forward spread.               

Let $(L_t )_{t \geq 0 } $ be the loss-given-default or LGD process, with $L_t $ the LGD if default happens at $t . $ In some of our results below $L_t $ is allowed to be stochastic, in which case we will if necessary assume it is adapted to the relevant filtrations generated by background processes such as the interest rate process.    
We assume that, for all $t > 0 $, $L_t $ is non-negative, and strictly positive with a non-zero probability: $\mathbb{E }(L_t ) > 0 $, where the expectation can be taken with respect to the objective probability measure, or an equivalent risk-neutral probability, if one exists. We will sometimes limit ourselves to LGDs which are known in advance, though possibly time-dependent, in which case we will write $L(t) $ instead of $L_t . $ 
Finally, we let $(r_t )_{t \geq 0 } $ be the risk-free short-rate process. 

Our first result generalizes the absence-of-arbitrage condition of the previous section to the continuous-time setting, under condition of non-negative short-rates.   
      
\begin{theorem} \label{thm:NAC_1} {Suppose that the risk-free short rate $r_t \geq 0 $ a.s. and that $T_0 < \tau \leq T_2 $ with positive (objective) probability. If there is no arbitrage in the CDS market, and if the protection fee is paid out continuously, then $(T - T_0 ) s(T_0 , T ) $ is a strictly increasing function of $T > T_0 $:
\begin{equation} \label{eq:NAC_1}
(T_2 - T_0 ) s(T_0 ,T_2 ) > (T_1 - T_0 ) s(T_0 , T_1 ) \ \mbox{if } T_2 > T_1 > T_0 .
\end{equation}   
The same inequality holds with discretely paid protection fees, provided the protection-payment dates of the 
two CDS contracts between $T_0 $ and $T_1 $ coincide. 
}   
\end{theorem}   
   
\noindent {\it Proof}. Recall that an arbitrage opportunity over the time-horizon $[0, T ] $ is a process $(X_t )_{t \geq 0 } $ representing a dynamic self-financing portfolio strategy which is bounded from below and has initial cost $X_0 = 0 $, while at $T $, $X_T \geq 0 $ a.s. and $X_T > 0 $ with strictly positive probability. Suppose now that (\ref{eq:NAC_1}) is not satisfied for some $T_1 < T_2 $:
\begin{equation} \label{eq:NAC_1a}
(T_2 - T_0 ) s_2 \leq (T_1 - T_0 ) s_1 ,   
\end{equation}
where we have written $s_1 := s(T_0 , T_1 ) $ and $s_2 := s(T_0 , T_2 ) $, to simplify notations. Then, necessarily,
$s_2 < s_1 . $ We set up a portfolio strategy at time 0 by shorting 
the CDS for $[T_0 , T_1 ] $ and going long the one for $[T_0 , T_2 ] $, and investing any cash earned into a savings account (as usual interpreted as borrowing if the amount is negative). This strategy costs 0 to set up. We verify that its value at $T_2 $ is non-negative with positive expectation.   
\medskip   
   
\noindent \textit{(i)} {\it Continuous protection payments}. If default happens before or at $T_0 $, both contracts are cancelled, and we receive nothing. If default happens before or at $T_1 $, the default payments of the two CDS contracts will cancel out, and the value of our position at $T_2 $ will be
$$
(s_1 - s_2 ) \int _{T_0 } ^{\tau } B_{t, T_2 } %e^{\int _t ^{T_2 } r_u du }   
dt > 0 ,
$$   
where    
$$   
B_{t, T } = e^{\int _t ^T r_u du }   
$$   
is the savings account. If default occurs between $T_1 $ (excluded) and $T_2 $ (included), we receive
 
\begin{eqnarray*}
&&(s_1 - s_2 ) \int _{T_0 } ^{T_1 } B_{t, T_2 } dt - s_2 \int _{T_1 } ^{\tau } B_{t, T_2 } dt + L_{\tau } B_{\tau , T_2 }  \\
&\geq & (s_1 - s_2 ) \int _{T_0 } ^{T_1 } B_{t, T_2 }  dt - s_2 \int _{T_1 } ^{T_2 } B_{t, T_2 } dt  + L_{\tau } B_{\tau , T_2 } \\
&\geq & \left( (s_1 - s_2 ) (T_1 - T_0 ) - s_2 (T_2 - T_1 ) \right) B_{T_1 , T_2 } %e^{\int _{T_1 } ^{T_2 } r_u du }   
+ L_{\tau } B_{\tau , T_2 } %e^{\int _{\tau } ^{T_2 } r_u du }   
\\
&=& \left( s_1 (T_1 - T_0 ) - s_2 (T_2 - T_0 ) \right) B_{T_1 , T_2 } + L_{\tau } B_{\tau , T_2 } \\
&\geq & L_{\tau } B_{\tau , T_2 } , %e^{\int _{\tau } ^{T_2 } r_u du } ,
\end{eqnarray*}
assuming (\ref{eq:NAC_1a}), where we used that the risk-free rate is non-negative for the third line, so that $B_{t, T_2 } \geq B_{T_1, T_2 } $ if $t \leq T_2 $ and $B_{t, T_2 } \leq B_{T_1 , T_2 } $ if $T_1 \leq t \leq T_2 . $ Finally, if $\tau > T_2 $, we receive the next to last line minus the LGD-term, which again is a non-negative amount. Since default will occur in $(T_0 , T_2 ] $ with a positive probability, our gain at $T_2 $ is non-negative and strictly positive with a positive probability, and since the value of our position is clearly bounded from below at intermediary times (it is in fact non-negative at all times by essentially the same arguments), we have an arbitrage opportunity. Note that our strategy is in fact a static one.   
\medskip   
   
\noindent \textit{(ii)} \textit{Discrete protection payments}.
Let $T_0 < t_1 < \cdots < t_{n_1 } = T_1 $ be the payment dates for the CDS contract over $(T_0 , T_1 ] $,   
which is augmented by payment dates $T_1 < t_{n_1 + 1 } < \cdots < t_{n_2 } = T_2 $ for the CDS over $(T_0 , T_2 ] . $ If default has not yet happened at $t_i $, then the protection fee for the sub-period $(t_i, t_{i + 1 } ] $ is payed at $t_{i + 1 } . $ If $s $ is the CDS-spread, his protection fee equals to $s (t_{i + 1 } - t_i ) $ if default has not yet occurred before $t_{i + 1 } $, or $s (\tau - t_i ) $ if default occurred after $t_i $ but before or at $t_{i + 1 } . $ This can be summarized by the formula   
$$   
s ( \tau \wedge t_{i + 1 } - \tau \wedge t_i ) ,   
$$   
where $a \wedge b := \min (a, b ) . $ 
With this notation, the value of our strategy at $T_2 $ is   
  
$$   
(s_1 - s_2 ) \sum _{i = 1 }^{n_1 - 1 } ( \tau \wedge t_{i + 1 } - \tau \wedge t_i ) B_{t_{i + 1 } , T_2 } - s_2 \sum _{i = n_1 } ^{n_2 - 1 } ( \tau \wedge t_{i + 1 } - \tau \wedge t_i ) B_{t_{i + 1 } , T_2 } + L_{\tau } B_{\tau , T_2 } .    
$$   
This value is clearly positive if $\tau \leq T_1 $, since $s_1 > s_2 $, while if $T_1 < \tau \leq T_2 $ , it is bounded from below by 

\begin{eqnarray*}    
&&(s_1 - s_2 ) \sum _{i = 1 } ^{n_1 } (t_{i + 1 } - t_i ) B_{T_1 , T_2 } - s_2 \sum _{i = n_1 } ^{n_2 - 1 } (t_{i + 1 } - t_i )B_{T_1, T_2} +L_{\tau } B_{\tau , T_2 } \\   
&=& B_{T_1 , T_2 } [(s_1 - s_2 ) (T_1 - T_0 ) - s_2 (T_2 - T_1 )] + L_{\tau } B_{\tau , T_2 } \\   
&=& B_{T_1 , T_2 } [s_1 (T_1 - T_0 ) - s_2 (T_2 - T_0 )] + L_{\tau } B_{\tau , T_2 } .   
\end{eqnarray*}   
where we used that $B_{t_{i + 1 } , T_2 } \geq B_{T_1 , T_2 } $ if $i < n_1 $ and $B_{t_{i + 1 } , T_2 } \leq B_{T_1 , T_2 } $ otherwise. We therefore have an arbitrage if inequality (\ref{eq:NAC_1}) is violated.

\hfill $\square $   
   
\begin{remark} \rm{The theorem remains valid if, in the discrete case, protection payments are made at the time of default instead of at the end of the protection period in which default falls: this simply amounts to replacing $B_{t_{i + 1 } , T_2 } $ by $B_{t_{i + 1 } \wedge \tau , T_2 } $ in the proof above, which can be bounded from above or below as before.     
   
Also, we can slightly weaken our assumption on the payment dates of the two CDS contracts, by supposing that the first contract has a set of payment dates $\{ T_0 < u_1 < u_2 < \cdots < u_{n_1 } = T_1 \} $ which contain the payment dates $\{ T_0 < T_1 < \cdots < t_{n_1 } < T_1 \} $ of the second contract with spread $s_2 $ for that period, while $T_1 $ is a payment date for the second contract.

Indeed, suppose that $T_1 > t_i = u_k < u_{k + 1 } < \cdots < u_{\ell - 1 } < u_{\ell } = t_{i + 1 } $ and that $\tau > t_i . $ If no default occurs between $t_i $ and $t_{i + 1 } $, then the value at $T_2 $ of our net protection payment for the period $(t_i , t_{i + 1 } ] $ is     
\begin{eqnarray*}   
&&s_1 \sum _{j = k } ^{\ell - 1 } (u_{j + 1 } - u_j ) B_{u_{j + 1 } , T_2 } - s_2 (t_{i + 1 } - t_i ) B_{t_{i + 1 } , T_2 } \\   
&\geq & s_1 \left( \sum _{j = k } ^{\ell - 1 } (u_{j + 1 } - u_j ) \right) B_{u_{\ell } , T_2 } - s_2 (t_{i + 1 } - t_i ) B_{t_{i + 1 } , T_2 } \\   
&=& (s_1 - s_2 ) (t_{i + 1 } - t_i ) B_{t_{i + 1 } , T_2 } ,   
\end{eqnarray*}   
while if the time of default $\tau $ falls within this period, a similar analysis shows that this net value is bounded from below by   
$$   
(s_1 - s_2 ) (\tau - t_i ) B_{t_{i + 1 } , T_2 } .   
$$   
Arguing as before, we find that we have an arbitrage opportunity if (\ref{eq:NAC_1}) does not hold.   
}   
\end{remark}

\section{Absence-of-Arbitrage conditions between CDS-rates and non-defaultable bond prices}\label{sectionAllRates} 

The proof of Theorem \ref{thm:NAC_1} above is completely elementary. We will now give a second proof which uses the fundamental theorem of asset pricing, and which will also allow us to generalize this theorem to the case where the risk-free rate can become negative. We will do so under a stronger assumption than Absence-of-Arbitrage, namely the existence of an equivalent martingale measure (or risk-neutral pricing measure) $\mathbb{Q } $ with respect to which the time-of-default $\tau $ and the risk-free short-rate process $(r_t )_{t \geq 0 } $ are independent. In other words, we assume a reduced-form credit risk model. An equivalent martingale measure exists assuming a suitably strengthened form of Absence-of-Arbitrage, such as Delbaen and Schachermayer's NFLVR condition (1994, \cite{Delbaen}). 
We will moreover assume, for a start that the protection is paid out continuously, at a rate of $s_0 (T_0, T ) $, and make some remarks on the case of discrete protection payments at the end of this section.   
\medskip   
   
Denoting the risk-neutral survival probability by\footnote{\bleu In a dynamic setting the spread as determined at $t > 0 $ would be written as $s_t (T_0 , T ) $ and the correct survival probabilities to use would be the conditional one, $q_t (u ) = \mathbb{Q } (\tau > u | \tau > t ) . $ \noir }
\begin{equation}
q(t ) := %q_0 (t) :=   
\mathbb{Q }(\tau > t ),   
\end{equation}   
the fair CDS spread at time 0 can these assumptions be expressed as
\begin{equation} \label{eq:CDS_spread_1}
s(T) := s_0 (T_0 , T ) = \frac{\mathbb{E }_{\mathbb{Q } } \left( L_{\tau } e^{- \int _0 ^{\tau } r_u du } \mathbf{1 }_{T_0 < \tau \leq T } \right) }{\int _{T_0 } ^T P_{0, t } q(t) dt } ,
\end{equation}
where $P_{0, t } := \mathbb{E }_{\mathbb{Q } } \left( e^{- \int _0 ^t r_u du } \right) $ is the price at time 0 of a non-defaultable 0-coupon bond maturing at $t . $

\begin{remark} \label{remark:LGDs} \rm{If $L_t =: L(t) $ is deterministic, the numerator equals   
\begin{equation} \label{eq:CDS_spread_2}
- \int _{T_0 } ^T L(t) P_{0, t } dq (t) ,
\end{equation}
This formula will still hold, with $L(t) := \mathbb{E }_{\mathbb{Q } } (L_t ) $ if the LGD process is also independent of $(r_t )_t $, 
For actual LGDs, such as those turning up in CVA calculations, the LGD is related to the present value, at time-of-default, of future discounted cash-flows, and would therefore in general depend on the short-rate process. If the CDS-contract sets the LGD is as a fixed fraction of a given nominal, it is clearly constant.   
}   
\end{remark}

To obtain a variant of Theorem \ref{thm:NAC_1} which holds if risk-free rates can be negative, it now suffices to replace $(T - T_0 ) $ by what we will call the {\it standardized (continuous-time) risk-free annuity} $A_0 (T_0, T ) $over $[T_0 , T ] $, defined as\footnote{An annuity would pay a constant interest of $c $ over a period $[T_0 , T ] $; if $c = 1 $, its present value at $t = 0 $ is given by (\ref{eq:Annuity}). }   
\begin{equation} \label{eq:Annuity}
A_0 (T_0 , T ) := \int _{T_0 } ^T P_{0, t } dt .   
\end{equation}

\begin{theorem} \label{thm:NAC_2} {\it Assume an equivalent martingale measure exists for the CDS market with respect to which $\tau $ %, $(L_t )_{t \geq 0 } $   
and $(r_t )_{t \geq 0 } $ are independent. 
Then $A_0 (T_0 , T ) s_0 (T_0 , T ) $ is an increasing function of $T . $ 
}   
\end{theorem}

\noindent {\it Proof}. Writing $A(T) $ and $s(T) $ for $A_0 (T_0 , T ) $ and $s_0 (T_0 , T ) $, and also introduce   
$$
A^d (T) := A^d _0 (T_0 , T ) := \int _{T_0 } ^T \, P_{0 , t } q(t) \, dt
$$
which is what we might call a defaultable standardized annuity. Then clearly   
$$   
A^d (T) s(T) = \mathbb{E }_{\mathbb{Q } } \left( L_{\tau } e^{- \int _0 ^{\tau } r_u du } \mathbf{1 }_{T_0 < \tau \leq T } \right)   
$$   
is increasing in $T $, since $\mathbf{1 }_{T_0 < \tau \leq T } $ is, so it suffices to show that $A (T) / A^d (T ) $ is increasing. But

$$   
\frac{\partial }{\partial T } \log \left( A(T) / A^d (T) \right) = \frac{P_{0, T } }{A(T) } %+ \frac{- L_T P_{0, T } q'(T) }{\mathbb{E }_{\mathbb{Q } } \left( L_{\tau } e^{- \int _0 ^{\tau } r_u du } \mathbf{1 }_{T_0 < \tau \leq T } \right) }   
- \frac{q(T) P_{0, T } }{\int _{T_0 } ^T P_{0, t } q(t) dt } \geq 0 ,   
%&\geq & \frac{- L_T P_{0, T } q'(T) }{\mathbb{E }_{\mathbb{Q } } \left( L_{\tau } e^{- \int _0 ^{\tau } r_u du } \mathbf{1 }_{T_0 < \tau \leq T } \right) } > 0 ,   
%\end{eqnarray*}   
$$   
since $q(t) $ is decreasing, and therefore   
$$
\int _{T_0 } ^T P_{0, t } \, q(t) dt \geq q(T) \int _{T_0 } ^T P_{0 , t } dt = q(T) A(T) .
$$
\hfill $\square $

\begin{remark} \rm{Theorem \ref{thm:NAC_2} implies Theorem \ref{thm:NAC_1} (modulo the stronger independence assumptions of the former), since
$$
\frac{T - T_0 }{A_0 (T_0 , T ) }
$$
is an increasing function of $T $ if %$P_{0, t } $ is   
the short-rate is non-negative a.s., since $P_0 (t) $ is then decreasing in $t \geq 0 $: this can be shown similarly by verifying that the derivative of the logarithm is non-negative. 
Note that in this case, 
the conclusion of Theorem \ref{thm:NAC_2} is stronger than the one of Theorem \ref{thm:NAC_1} (see also the graphical  interpretation of these Absence-of-Arbitrage relations below). An advantage however of (\ref{eq:NAC_1}) is that it is formulated purely in terms of the CDS spreads,  without reference to the risk-free yield curve.
\medskip

}   
\end{remark}   
   
\subsection{A No-Arbitrage relation between CDS and IRS rates} We can reformulate Theorem \ref{thm:NAC_2} as a no-arbitrage relation in terms (continuously paid) Interest Rate Swaps (IRS) rates. Recall that the fair (forward) rate for a continuously exchanged fixed-for-floating IRS over $[T_0 , T ] $ is given by%\footnote{\bleu use lower case \noir $i_0 (T_0, T ) $ \bleu as notation for IRS rates? \noir }
\begin{equation} \label{eq:IRS}
I_0 (T_0 , T ) = \frac{P_{0, T_0 } - P_{0, T } }{A_0 (T_0 , T ) } ,   
\end{equation}   
with $I_0 (T_0, T_0 ) := \lim _{T \to T_0 + } I_0 (T_0 , T ) =  - P_{0, T_0 } ^{-1 } \partial P_{0, T } / \partial T |_{T = T_0 } $, the $T_0 $-forward rate at time 0, where we used (\ref{eq:IRS}).   
\medskip   
   
Theorem \ref{thm:NAC_2} then implies that $\left( P_{0, T_0 } - P_{0, T } \right) s_0 (T_0 , T ) / I_0 (T_0 , T ) $ is an increasing function of $T > T_0 . $ Equivalently, if we introduce the $T_0 $-forward price $F_0 (T_0 , T ) := P_{0, T } / P_{0, T_0 } $, then we have the   
   
\begin{corollary} {\it %Assuming an EMM exists,   
Under the assumptions of theorem \ref{thm:NAC_2},   
\begin{equation} \label{eq:NAC_2bis}
\left( 1 - F_0 (T_0 , T ) \right) \frac{s_0 (T_0 , T ) }{I_0 (T_0 , T ) }
\end{equation}
is increasing in $T \geq T_0 . $   
}   
\end{corollary}   
   
One can go one step further, and express $F_0 (T_0 , T ) $ in terms of the IRS-forward rates: see Appendix A.
This however involves a double integration of IRS-rates with respect to their maturity variables, and the corollary suffices in practice, since we can observe forward bond-prices along with forward IRS rates.   
   
\subsection{%Geometrical   
Graphical interpretation of the No-Arbitrage relations}   
   
The No-arbitrage condition of Theorem \ref{thm:NAC_1} has a simple graphical interpretation: suppose we have a collection of observed rates $s_i = s(T_0 , T_i ) $ and that we plot the points $s_i $ against $x_i := T_i - T_0 $ in the $(x, s ) $-plane. Take a point $(x_a , s_a ) $ and draw the hyperbola $s = C / x_a $ which passes through this point (so that $C = x_a \cdot s_a $). Then all points $(x_i, s_i ) $ for which $x_i \geq x_a $ should lie above this hyperbola, if not, there is an opportunity for arbitrage. 

\begin{figure}
\caption{Geometric Interpretation for Arbitrage Condition - Theorem 1}\label{figureTheorem1} 
\centering  
\hspace*{-1cm}  
\includegraphics[scale=.62]{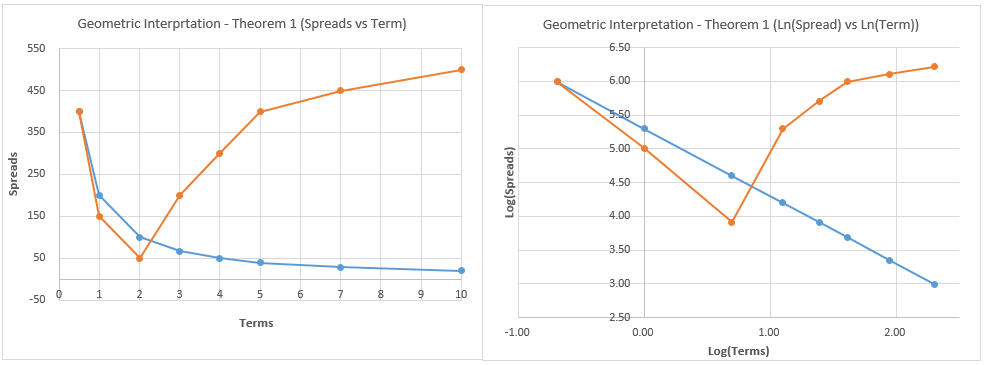} 
\end{figure}

In Figure \ref{figureTheorem1}, a CDS curve highlighted in orange color is plotted against a hyperbola going through the $(x_a, s_i)$ where $x_a=0.5$ and $s_a=400$ basis points. Clearly, together with $s_a$, any points $s_i$ located below the hyperbola can form a paired trade to construct arbitrage opportunity as in the proof of Theorem \ref{thm:NAC_1}, which makes it easy to detect such arbitrage opportunities at a glance.   
   
Alternatively, one can do a log-log plot: $\log s_i $ against $\log x_i $, in which case the points $(\log x_i , \log s_i ) $ for which $\log x_i > \log a_i $ should lie above the straight line with slope -1 which passes through $(\log x_a , \log s_a ). $ The right panel of Figure \ref{figureTheorem1} illustrates the graphical interpretation for Theorem \ref{thm:NAC_1} in log-log scale.
    
\medskip   

A similar construction applies for Theorem \ref{thm:NAC_2}: it suffices to plot $s_i $ against $A_i := A_0 (T_0 , T_i ) $ respectively $\log s_i $ against $\log A_i . $

\subsection{Discrete protection payments}

\medskip   
   
In practice, the protection leg of a CDS is paid discretely instead of continuously. If we consider families of co-initial CDS-contracts with maturities and protection payment dates coming from a discrete set of tenor-dates $T_0 < T_1 < \cdots < T_n < \cdots $, then the protection buyer of a CDS contract whose maturity is $T_n $ pays $s (T_i \wedge \tau - T_{i - 1 } \wedge \tau ) $ at dates $T_i $ for $i = 1 , \ldots ,  n $: $s (T_i - T_{i - 1 } ) $ if default hasn't occurred yet, and $s (\tau - T_{i - 1 } ) $ if $T_{i - 1 }< \tau \leq T_i . $ One can repeat the analysis to arrive at the following result:   
  
let 
$$   
A_n := \left( \sum _{i = 1} ^n (T_i - T_{i - 1 } ) P_{0 , T_i } \right) ,   
$$   
be the discrete-time standardized annuity, and let $s_n := s(T_0 , T_n ) $ be the fair CDS spread. Then   
   
\begin{theorem} If an equivalent martingale measure exists and if $\tau $ is independent of the risk-free interest rate process, then $A_n s_n $ is an increasing function of $n . $   
\end{theorem}   
   
\noindent {\it Proof}. The fair spread now equals   
$$  
s_n %:= s(T_0, T_n ) :=   
= \frac{\mathbb{E }_{\mathbb{Q } } \left( L_{\tau } e^{- \int _0 ^{\tau } r_u du } \mathbf{1 }_{T_0 < \tau \leq T } \right) }{A^d _n } ,   
$$   
where   
$$   
A^d _n := \sum _{i = 1 } ^n \mathbb{E }_{\mathbb{Q } } (T_i \wedge \tau ) - \mathbb{E }_{\mathbb{Q } } (T_{i - 1 } \wedge \tau ) P_{0, T_i } ,   
$$   
%Letting $A_n := \sum _{i = 1} ^n (T_i - T_{i - 1 } ) P_{0 , T_i } $,   
and it again suffices to verify that $A_n / A^d _n $ is increasing in $n . $ Now an easy computation shows that $\mathbb{E }_{\mathbb{Q } } (T \wedge \tau ) = Q (T ) $, where   
$$   
Q(T ) := \int _0 ^T q(u) du .   
$$   
It follows that $(Q (T_i ) - Q(T_{i - 1 } ) ) / (T_i - T_{i - 1 } ) $ is decreasing in $i $, since   
$$   
q(T_i ) (T_i - T_{i - 1 } ) \leq \int _{T_{i - 1 } } ^{T_i } q(u ) du = Q(T_i ) - Q(T_{i - 1 } ) \leq q(T_{i - 1 } ) (T_i - T_{i - 1 } ) ,   
$$   
and therefore 
$$   
\frac{Q (T_{i + 1 } ) - Q (T_i ) }{T_{i + 1 } - T_i } \leq q(T_i ) \leq \frac{Q (T_i ) - Q(T_{i - 1 } ) }{T_i - T_{i - 1 } } .   
$$   
Now   
$$   
\frac{A_{n + 1 } }{A^d _{n + 1 } } - \frac{A_n }{A^d _n } = \frac{\sum _{i = 1 } ^n \left( (T_{n + 1 } - T_n ) (Q(T_i ) - Q(T_{i - 1 } ) - (T_i - T_{i - 1 } ) (Q(T_{n + 1 } - Q(T_n ) ) \right) P_{0, T_i } P_{0, T_{n + 1 } } }{A^d _{n + 1 } A^d _n } ,   
$$   
and each term in the sum in the numerator is positive since $(Q(T_i ) - Q(T_{i - 1 } ) ) / (T_i - T_{i - 1 } ) \geq (Q(T_{n + 1 } ) - Q(T_n ) ) / (T_{n + 1 } - T_n ) $ for $i \leq n . $   
\medskip   
   
One checks by a similar computation that if interest rates are non-negative, and $P_{0 , T_n } $ therefore is decreasing in $n $, then $(T_n - T_0 ) / A_n $ divided by the numerator is increasing in $n $, so that we find as in section 2 that $(T_n - T_0 ) s_n $ is increasing.

\section{Further Empirical Evidence}\label{sectionEmpirics}
 
Economists often cite inverted yield curves as signals for a pending economic crisis. However, inversions for CDS-curves does seem not to have been systematically investigated in the literature. In this section we present some further evidence for CDS anomalies summarized across the sample period of our CDS data. We consider the implications of such anomalies for conditional default probabilities.     

\subsection{CDS-term-structure anomalies across time}\label{sectionFurtherAnomalies}
    
In section \ref{sectionAnomalies}, we presented a cross-sectional overview of the CDS-term-structure arbitrages (or anomalies) which fail the No-arbitrage condition of Theorem \ref{thm:NAC_1}.   
 
We now take a further look at these anomalies across the same sample period using the same CDS data set, where we consider contracts with $T_0 = 0 $, that is, starting on the day they are quoted.  It is convenient to introduce the {\it Maturity Adjusted Spread Ratio} or MAR, defined as   
$$   
MAR := T_1 s(T_1 ) / T_2 s(T_2 ) ,   
$$   
so that Theorem \ref{thm:NAC_1} there is arbitrage for the pair of traded if $MAR > 1 . $ Equivalently, there is arbitrage if the {\it slope} $s(T_1 ) / s(T_2 ) $ is bigger than or equal to the maturity ratio $T_2 / T_1 . $ The MAR is a metric for the \textit{strength} of such an arbitrage opportunity.   
\smallskip

\medskip
   
Figure \ref{figureSummaryofAnomaliesAcrossTime} graphs the total number of anomalies per month or Anomaly Counts in our dataset. Figure \ref{figureSummaryofCDSArbitrage} graphs these in red for each of the four maturity pairs we examined, together with the MARs in blue to indicate the strength for arbitrage opportunities based on each maturity pair.

\begin{figure}
\caption{Arbitrageability/MARs across strategies and rating groups}\label{figureArbitrageability} 
\centering    
\includegraphics[scale=.7]{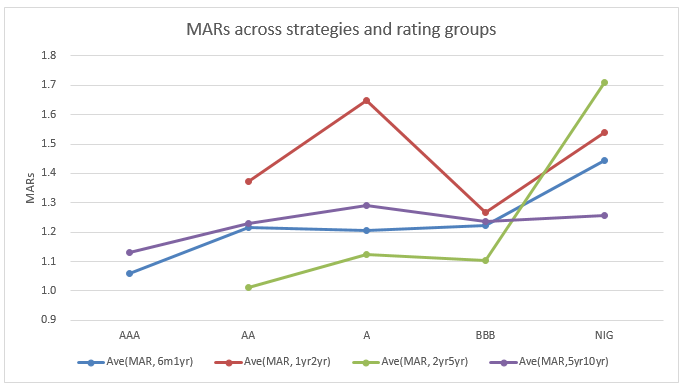} 
\end{figure}

On the basis of these two figures
and on Figure \ref{figureArbitrageability} and Table \ref{tableArbitrageability}, we can make the following observations:   

\begin{enumerate}

\item We can distinguish three different phases in the anomalies count of Figure \ref{figureSummaryofAnomaliesAcrossTime} which correspond with three phases of the credit crunch: 
% those for the crisis based on monthly counts of anomalies:   
(\textit{i})  a \textit{build-up phase}, from July/August-2007 (when BNP Paribas terminated investors' withdrawal from three of the hedge funds it administered, often considered as the start of the crisis) to March-2008 (when US investment bank Bear Stern collapsed), coinciding with the initial phase of the financial crisis;   
% before the liquidity crunch after  Lehman's bankruptcy;   
(\textit{ii}) a \textit{liquidity dry-up phase}, from April-2008 to May-2009, during which the number of anomalies decreased while the market was hit by the liquidity crunch; \textit{(iii)} a \textit{post-crisis liquidity recovery and market  uncertainty phase}, from June-2009 (the official ending of the crisis) onwards during which, as funding and market liquidity recovered, the number of anomalies increased as the market searched for directions amongst ongoing uncertainty, in part related to regulatory reforms hanging over the financial markets. According to Battalio and Schultz (2011, \cite{Battalio}), regulatory confusion and uncertainty created significant bid-ask spreads in the equity option market, and  the CDS market was not immune from this either.        
\begin{figure}
\vspace{-2mm}
\caption{The number of Anomalies across time}\label{figureSummaryofAnomaliesAcrossTime} 
\centering  
\includegraphics[scale=.62]{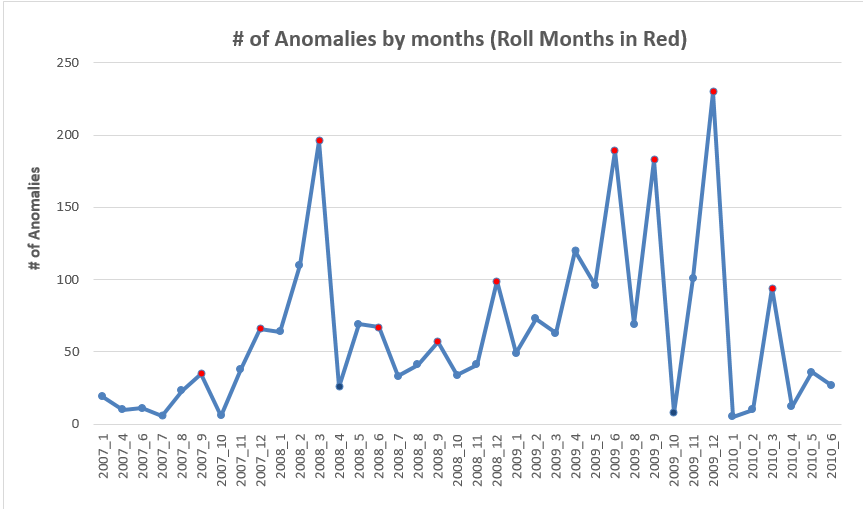} 
\end{figure}

\item Figure \ref{figureSummaryofAnomaliesAcrossTime} shows that 8 out of the 12 peak-months (highlighted in Red) coincide with the "Roll Months", which are the months containing so-called Roll Dates\footnote{Our CDS contracts are of 1999/2003 ISDA formats; Roll Dates refer to the 20th of March, June, September and December when less liquid off-the-run CDS contracts expire and more liquid on-the-run ones are traded.}. During the Roll Months, the increased number of anomalies can at least partly be attributed to an increased number of expiring and lack of liquid CDS contracts before the Roll Dates.

\begin{figure}
\caption{Anomaly Counts and Arbitrage Opportunity Strengths for Four Strategies}\label{figureSummaryofCDSArbitrage} 
\centering  
\includegraphics[scale=.63]{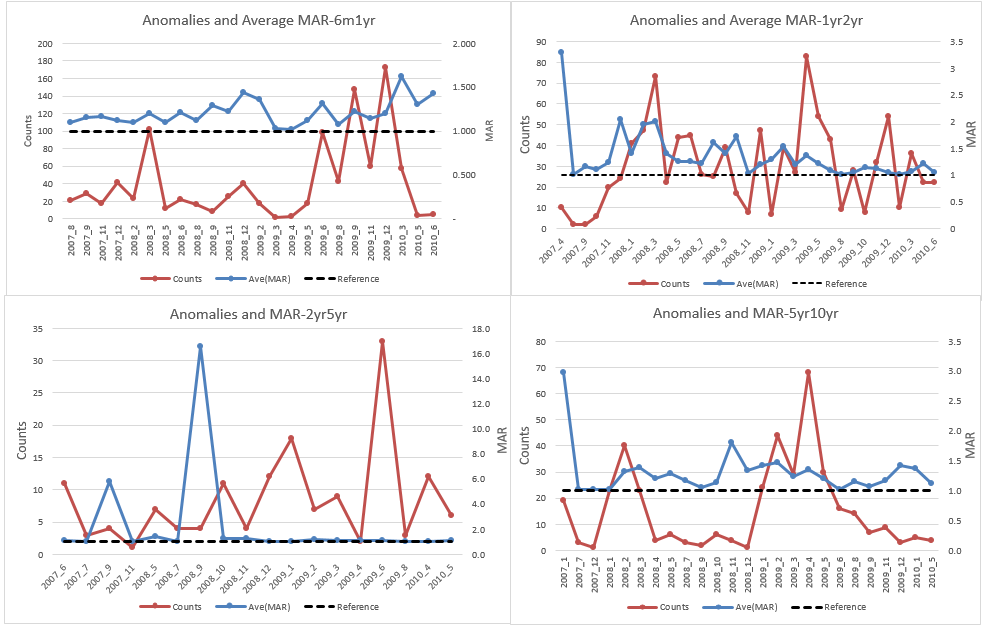} 
\end{figure}

\item Across the four %panels indicating four   
different arbitrage strategies of Figure \ref{figureSummaryofCDSArbitrage}, the '6m-1yr' maturity pair had the highest number of arbitrages, 987,  %or anomalous trades,   
closely followed by 975 for the '1yr-2yr' pair. For the '2yr-5yr' pair we found 151 arbitrage opportunities and for the '5yr-10yr' pair, 388. In our data set, CDS-curve arbitrage opportunities occurred most frequently at the short-end of the curve and less frequently for the longer maturity contracts, and the least for 5-year term contracts which are the most liquidly traded. These observations are  consistent with an economic principle: when a name gets stressed, its CDS curve often gets inverted at the short-end (such as the 6-month maturity) of the curve and tends to go back to be normally upward-sloping after that. This is because investors tend to believe that if a stressed name survives in the short-term, it will bounce back over longer terms.       
   
\item Figure \ref{figureSummaryofCDSArbitrage} also depicts     
the MARs in contrast with the reference line ($y=1$ indicated by a dark dotted line); as explained above, the greater the (monthly average) MAR number exceeds the reference line, the stronger or the more arbitrage-able the arbitrage opportunity can be. Also, MARs tend to be less volatile than the anomaly counts, except for some outliers. For example, the spike of 16.53 in the MARs shown in Table \ref{tableArbitrageability} for the  '2yr-5yr' couple is caused by Freddie Mac (a US home insurance company) because its 2-year CDS spread went up to 1538 basis points after Lehman Brother's bankruptcy while its other spreads  
stayed normal during the month of Sept-2008, as the market expected Freddie Mac to survive after the short-run stress. (The market turned out to be correct: Freddie Mac survived after receiving a governmental bailout.) 

\item In Table \ref{tableArbitrageability}, we present the average MAR of the anomalies for the four maturity pairs split out by rating class. We excluded one anomalous trade related to Freddie Mac as an outlier from the '1yr-2yr' and '2yr-5yr' pairs because it was rated AAA even as it received the US government's bailout. Figure \ref{figureArbitrageability} and Table \ref{tableArbitrageability} show that the %steepness   
MAR tends to increase when we go from higher credit-quality ratings groups (AAA) to lower ones ( BBB or Non-investment-Grade rated group (or NIG, ratings lower than), with the exception of the A-rated group for all four strategies. This may be related to an observation of Biswas, Nikolova and Stahel (2015, \cite{Biswas}), who found that CDS contracts on lower rated entities commanded a higher liquidity premium. Such  liquidity premia would disincentivise would-be arbitrageurs to take advance of these arbitrage opportunities since it would make it costly to close out the arbitrage strategy prematurely should they wish to do so, explaining why such opportunities are tolerated by the market.

\begin{table}

\caption{Arbitrageability/MARs across strategies and rating groups: numerical results}\label{tableArbitrageability} 
\centering   
\includegraphics[scale=.75]{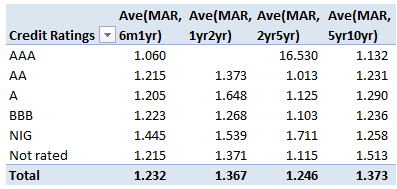} 
\end{table}

\end{enumerate}

\subsection{Numerical example of arbitrage profits across time: Microsoft 5yr-10yr strategy}
\begin{figure}

\caption{CDS-Curve arbitrage Case Study: Microsoft (5yr10yr Strategy)}\label{figureMSFTCDSPnL} 
\centering  
\hspace*{-1cm}  
\includegraphics[scale=.8]{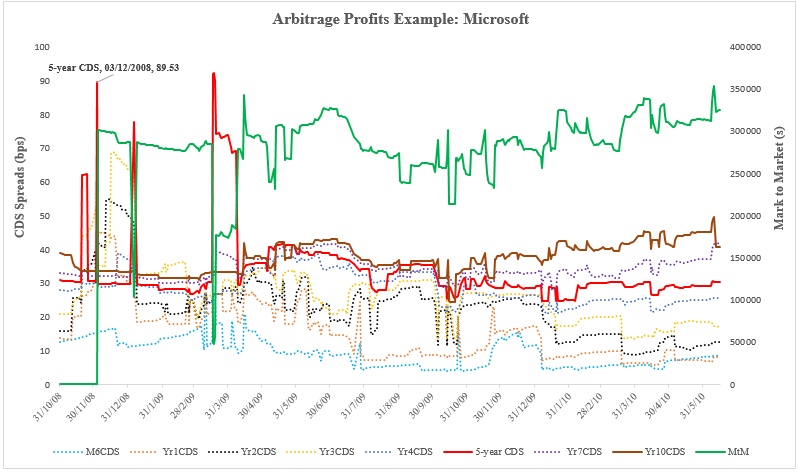} 
\end{figure}

If two CDS contracts of different maturities, e.g., the 5-year and 10-year, are quoted on the market with spreads $s_{5yr}, s_{10yr} $ which violate the No-arbitrage condition of Theorem (\ref{thm:NAC_1}), we can construct an arbitrage strategy %as $H(\theta_1, \theta_2):=(-s_{5yr}, s_{10yr})$, i.e.,   
by selling the 5-year default protection and at the same time buying the 10-year default protection. %based on the CDS market quotes for the two contracts.   
Observing the market quotes for Microsoft on 03/12/2008, we form such a strategy and calculate the Mark-to-Market value for two CDS contracts with notional amounts of 10 Million dollar each. Figure \ref{figureMSFTCDSPnL} presents the computed Mark-to-Market values (graphed in Green from the right $y$-axis) from ISDA's standard CDS model\footnote{based on a reduced form model for standard CDS as the one we use in Section 3; see for example O'Kane, 2008, \url{http://www.cdsmodel.com/cdsmodel/}} of the resulting strategy using CDS market data from 03/12/2008 to 30/06/2010. We assume zero interest for the observation period, given that interest rates were extremely low then. Besides, as shown in Appendix \ref{sectionRiskSensitivities}, interest rate levels only have a limited impact on CDS  valuation.

At the inception of the strategy, the 5-year mid-CDS rate on 03/12/2008 is 89.53 basis points while the 10-year one is 33.55 basis points, $\frac{89.53}{33.55}=2.669 > \frac{10}{5} . $ The Mark-to-Market value is zero on 03/12/2008, but raises to a strictly positive value immediately afterwards, and consistently stays positive throughout the observation period.   
\medskip   
   
Besides the Mark-to-Market value of our arbitrage strategy, Figure \ref{figureMSFTCDSPnL} also plots CDS rates (right $y $-axis) for protection on Microsoft for eight different maturities, for each date between 31/10/2008 to 30/06/2010.  The CDS rates for the 5-year and 10-year maturities are represented by solid lines in red and brown and the other terms by dotted lines. Other spikes, where the 5-year CDS rate (in red) overshoots the 10-year rate (brown) can be observed later in the period. The 5-year rate is more volatile than the 10-year one, which is probably related to its greater liquidity. In particular, this indicates that arbitrage opportunities like these are not one-offish, instead, they can be recurrent.
  
\begin{figure}

\caption{Curve-arbitrage Case Study: Microsoft (5yr10yr Strategy)}\label{figureMSFTCDSArbitrage} 
\centering  
\includegraphics[scale=.75]{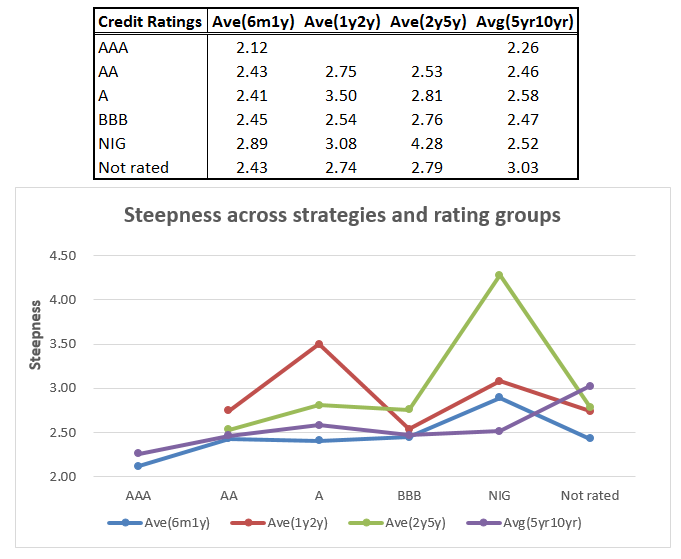} 
\end{figure}

\subsection{Implications of Arbitrage: Nonsensical Default Probabilities}

\begin{figure}
\caption{Negative Hazard Rates for AIB (6m1yr) and Microsoft (5yr10yr) Strategy}\label{figureNegativeHazardRate} 
\centering  
\includegraphics[scale=.45]{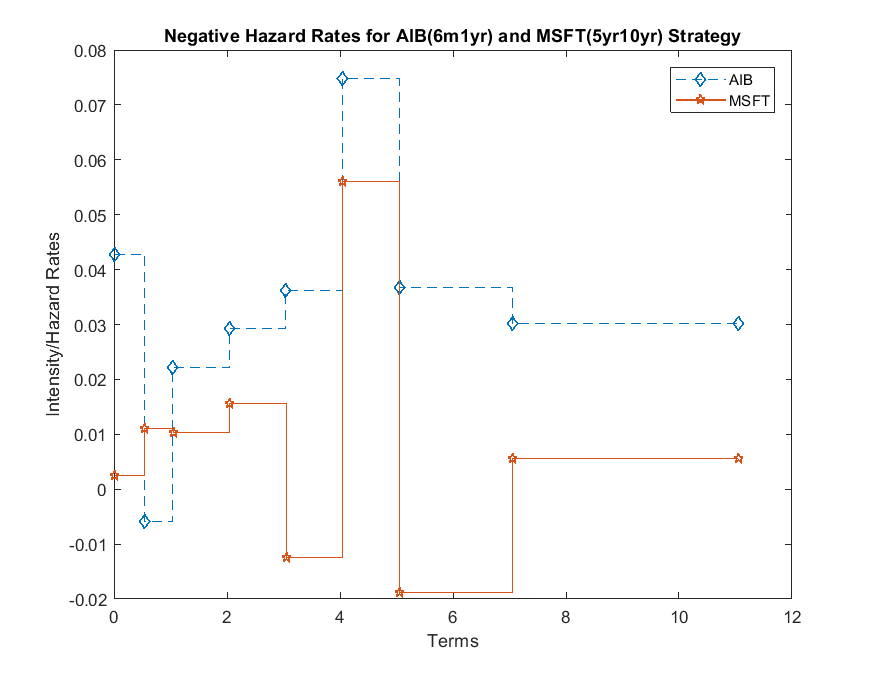} 
\end{figure}

As we already noted in section \ref{sectionNoarbitragecondition}, Remark \ref{remark1}, violation of the No-arbitrage condition of theorem \ref{thm:NAC_1} can lead to nonsensical conditional default probabilities as computed in a standard reduced-form model. Observing that the CDS term structures for AIB on 04/12/2008 and Microsoft on 03/12/2008 respectively do violate this the No-arbitrage condition %given in Theorem (\ref{thm:NAC_1}), we investigate to check for   
look at the implications for each entity's the conditional default probabilities, using a standard reduced-form model: the ISDA standard CDS model. This model bootstraps piecewise constant conditional default probabilities or hazard rates from a given CDS curve. Table \ref{tableNegativeHazardRate} shows that the bootstrapped hazard rates for 4-year and 7-year terms for Microsoft are negative; also, the bootstrapped 1-year hazard rate for AIB is negative (highlighted in red colors). In both cases, this violates the elementary definition of a probability. Figure \ref{figureNegativeHazardRate} depicts the same numerical results in a graph.      

\begin{table}
\caption{Numerical results of Hazard Rates for AIB (6m1yr) and Microsoft (5yr10yr) Strategy}\label{tableNegativeHazardRate} 
\centering  
\includegraphics[scale=0.75]{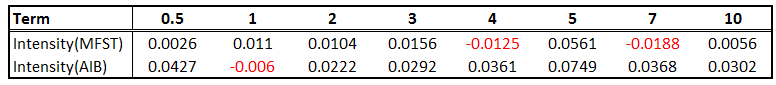} 
\end{table}   

\subsection{CDS Term Structure and Credit Valuation Adjustment}\label{sectionCDSTermStructureImpactsonCVADVA}   
  
We discuss the relation between CDS term structure arbitrage opportunities and CVA. In one direction, the risk-neutral hazard rates and default probabilities which are used for the computation of the CVA of a given OTC derivative contract are backed out from observed CDS curves. If these curves present arbitrages in the sense of our theorems 1-3, these hazard rates can become negative, which will result in an over-valuation of the CVA. Since CVAs are not traded, there are not likely to be other consequences.   
   
In the other direction, CVA costs can provide an explanation for why these arbitrage oportunities are not exploited. Indeed, the would-be arbitrageur faces two kinds of counterparty credit risks: that of the protection buyer of the smaller maturity CDS contract not paying all of his protection fees between $T_0 $ and $T_1 $ and that of the protection seller not paying the loss-given-default, in particular for default occuring before $T_1 $, when this payment should compensate the loss-given-default payment the arbitrageur himself should make. In other words, the CVA of the arbitrage strategy will make its initial value negative instead of 0. Unaccounted counterparty credit risks to be borne by a would-be arbitrageur can thus help to explain the existence of the significant and persistent amount of CDS-term-structure arbitrage in our study. We note in this respect that Brigo and Capponi (2014, \cite{BrigoCapponi}) show in an example that the impact of CVA on the Mark-to-Market value of a 5-year single-name CDS payer contract can vary up to 60 basis points, depending on the default correlation between the two parties.   

In addition, other types of valuation adjustments can also explain the arbitrage opportunities observed in our study in form of implicit costs and risks. For example, the counterparty credit risk run by the arbitrageur may be alleviated through the posting of collateral (e.g. in a context of centrally cleared contracts) but the arbitrageur himself would then have to post collateral also, which then corresponds to a non-zero initial investment. Furthermore, liquidity risks might prevent the arbitrageur to prematurely close-out his position. Biswas (2015, \cite{Biswas}) studied the liquidity risk component in a real-world setting, as measured by the bid-ask spread for CDS contracts, based on mainly post-crisis CDS market data. They found that the liquidity risk premiums for CDS contracts are between 12 and 14 basis points depending on whether the contracts are between a dealer and another dealer or between dealer and a buy-side investor. they also found that an entity with lower credit rating has to pay higher liquidity premiums. Since we use data from the time period leading to and during the financial crisis, we expect the liquidity premiums for our mainly dealer and buy-side investors (or end-users) trades to be higher than these numbers.

\section{Conclusions \& Research Directions} \label{sectionConclusion}

We derived criteria for absence of arbitrage in CDS term structures, under various assumptions (discrete or continuous protection payments, interest rate environments which are positive or not) with no or only minimal modelling assumptions. These criteria provide simple checks for arbitrage opportunities in the CDS market. Based on these criteria and using a comprehensive data set, we found that a significant amount of CDS-term-structure arbitrage opportunities existed in the CDS market prior and during the financial crises.   
  
Violation of the no-arbitrage conditions implies nonsensical conditional default probabilities which will impact on the calculation of CVA and DVA. The existence of such arbitrage opportunities is likely to be connected with hidden risks (counterparty credit risk, liquidity risk) and costs (collateral) which prevents them to be exploited, and which prior to the 2007-09 crises were not systematically taken into account in the pricing of OTC derivatives.

The progress being made in the theory and practice of the valuation adjustment should contribute to the gradual disappearance of these arbitrage opportunities, as should the increase in efficiency and transparency due to the standardisation of OTC derivative contracts and the use of modern information technology. Furthermore, central clearing should contribute to greater liquidity. The existence of arbitrage opportunities should therefore not discourage the financial community from continuing to use the CDS market for the calibration of counterparty default risks.   
\medskip   
   
As possible directions for future research we mention the expansion of our study to maturity pairs other than the four we considered here, and other data sets.  We also believe that our analysis can be extended to cover sovereign CDS while accounting for the exchange rate between the protection and premium legs in context of the so-called Quanto CDS; no study has been done on the term-structure arbitrage of such contracts. On a theoretical level, it would be interesting to characterise arbitrage-free CDS term structures: we mention here that one can construct examples of CDS curves which satisfy the criteria of theorems 1 and 2, but which nevertheless cannot be a CDS curve in an arbitrage-free model. This is work in progress.

\appendix

\section{Forward bond-prices and IRS rates}

\begin{proposition} \label{thm:IRS_to_bond} {\it For $T_0 \leq U \leq T $ let
\begin{equation} \label{eq:IRS_to_bond:1}
\Phi _0 (U, T ) := e^{- \int _U ^T I_0 (T_0 , V ) dV } .
\end{equation}
Then
\begin{equation} \label{eq:IRS_to_bond:2}
F_0 (T_0, T ) = 1 - I_0 (T_0, T ) \int _{T_0 } ^T \, \Phi _ 0 (U, T ) \, dU .
\end{equation}
In particular, the forward IRS curve uniquely determines the forward bond prices.   
}   
\end{proposition}   

\begin{corollary} \label{corr_NA_CDS_IRS} {\it If there is no arbitrage in the combined CDS - IRS market then (assuming the existence of an equivalent martingale measure)    
\begin{equation} \label{eq:NA_CDS_IRS}
T \to \left( \int_{T_0 } ^T \Phi _0 (U, T ) dU \right) s_0 (T_0 , T )
\end{equation}
is an increasing function on $(T_0 , \infty ) . $   
}   
\end{corollary}

\noindent {\it Proofs}. The corollary follows by inserting (\ref{eq:IRS_to_bond:2}) into (\ref{eq:NAC_2bis}). %As for the proof of theorem \ref{thm:IRS_to_bond}, this foreshadows a more general result for CDS curves which is to be established in the next section (and which in fact contains the present one as a special case). We nevertheless sketh the proof\footnote{\bleu perhaps to be skipped in the final version to the advantage of the more general result, to avoid too much duplication?\noir }.   
The proof of (\ref{eq:IRS_to_bond:2}) we set up and solve an ODE for the forward bond prices.  We fix $T_0 $, and write $I(T) $ for $I_0 (T_0, T ) $ and $F(T) $ for $F_0 (T_0 , T) . $ On dividing both sides of the fundamental relation (\ref{eq:IRS}) by $P_{0, T_0 } $, we can rewrite this as
$$
A^F (T) I(T) = 1 - F(T) ,
$$
where $A^F (T) := A^F _0 (T_0 , T ) := \int _{T_0 } ^T F(t) dt $ (which is nothing else but the forward price of $A_0 (T_0 , T ) $). Differentiation with respect to $T $ gives $- F'(T) = A^F (T) I' (T) + F(T) I(T) = (1 - F(T) ) I'(T) / I(T) + F(T) I(T) $, where we used the fundamental relation again, or
$$
F' + \left( I - \frac{I ' }{I } \right) F = - \frac{I'}{I } .
$$
This as a linear non-homogeneous ODE for $F(T) $ for $T > T_0 $ with initial condition $F(T_0 ) = 1 $, whose solution is given by (\ref{eq:IRS_to_bond:2}).   
  
\medskip   
   
We make some further observations:   
\begin{enumerate}   

\item In a  dynamic setting relations such as the ones established here hold between $s_t (T_0, T ) $ and/or $P_{t, T } $ and $I_t (T_0 , T ) $ and put restrictions on a joint dynamic CDS-IRS model which simultaneously tries to mode the swap and CDS rates.   
   
\item If interest rates (instantaneous rates or forward rates in an HJM model) are non-negative, then forward prices $F_0 (T_0, T ) $ are decreasing in $T $, and consequently $I_0 (T_0, T ) \int _{T_0 } ^T \Phi _0 (U, T ) dU $ has to be increasing in $T $: this puts a restriction on possible IRS forward curves at a given time 0. Similar conditions can be derived for CDS forward curves.   

\item In an HJM model, forward prices can be expressed in terms of instantaneous forward rates $f_{0, T } $ as
$$
F_0 (T_0 , T ) = e^{- \int _{T_0 } ^T f_{0, U } dU } .
$$
One can ask for a relationship between the $f_{0, T } $'s and the forward swap rates $I(T) = I_0 (T_0 , T ) . $ Recall that
$$
(1 - F ) (I' / I ) + F I = - F' = f F ,
$$
where $f(T) := f_{0, T } . $ Differentiating this relation once more, and then using it to eliminate the $1 - F $-term form the resulting expression, we find a first-order Ricatti equation for the forward rate $f(T) $, $T \geq T_0 $,
\begin{equation}
f' = f^2 + a f + b
\end{equation}
with $a = (I'' / I' ) - I = (\log I' )' - I $ and $b = - I'' I / I' = - I (\log I ) ' $ and initial condition $f(T_0 ) = I(T_0 ) := \lim _{T \to T_0 } I(T) . $ The solution of this Ricatti equation is of course implicitly already contained in  (\ref{eq:IRS_to_bond:2}), since the instantaneous forward rate is minus the logarithmic derivative of the forward: $f_{0, T } = - \partial _T F_0 (T_0 , T ) / F_0 (T_0 , T ) $ for $T > T_0 . $ However, in practice, one might also solve this ODE numerically.

\end{enumerate}

 \section{Risk Sensitivities for 5yr-10yr Microsoft Arbitrage Strategy }\label{sectionRiskSensitivities}

\begin{figure}
\vspace{-10mm}
\caption{DV01, CR01 and MtM for Microsoft (5yr10yr Strategy)}\label{figureMSFTDV01Arbitrage} 
\centering  
\includegraphics[scale=.5]{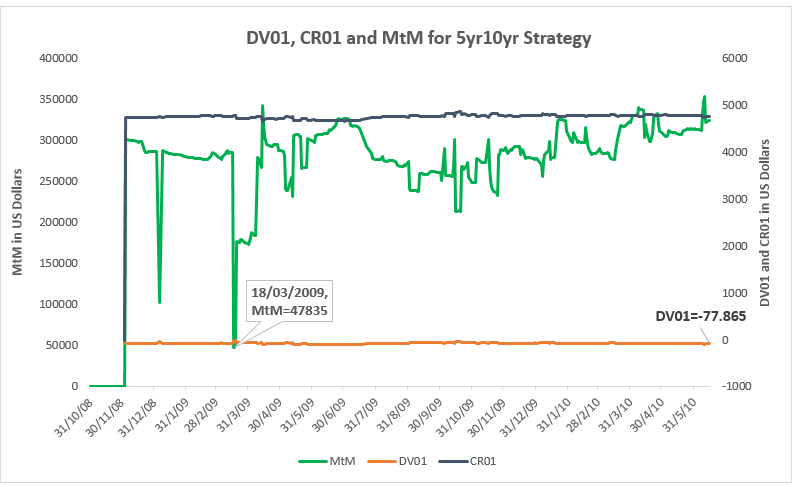} 
\end{figure}

Following O'Kane (2008), we measure the Interest Rate Risk Sensitivity by so-called DV01, which measures the Mark-to-Market value denoted by $V(r,s)$ changes with regard to 1 basis point parallel shift for interest rate level or $DV01=V(r+1,s)-V(r,s)$. Similarly, we measure Credit spread risk sensitivity by CR01 as the Mark-to-Market value changes with regard to 1 basis point parallel shift for CDS curve; i.e. $CR01=V(r,s+1)-V(r,s)$. We note that: (1) Figure \ref{figureMSFTDV01Arbitrage} displays the DV01, CR01 together with MtM results for the 5yr10yr Arbitrage strategy of Microsoft described above. (2) The DV01 for this strategy is between -103 and -11 dollars throughout the data period in our sample; whereas, the CR01 for this strategy is between 4,761 and 4,870 dollars for the sample period. Clearly, CR01 or credit risk instead of DV01 or interest rate risk is the main driver for the MtM for this strategy. (3) We note that, the above observation can be generalized to all single-name CDS;  i.e., interest rate sensitivity for single-name CDSs is much smaller than credit spread risk with regard to their respective rates. 

\begin{thebibliography}{9}

\bibitem{Baietal}
Bai J. and Collin-Dufresne P., November 2013, ''The CDS-Bond Basis'', \textit{AFA 2013 San Diego Meetings Paper}, Georgetown Univ., Dept. of Finance and Ecole Polytechnique Fédérale de Lausanne

\bibitem{BISstats}
Banks for Int'l Settlements, July 2017, \textit{Statistical release: OTC derivatives statistics at end June 2017}.

\bibitem{Battalio}
Battalio, R. and Schultz P., 2011, ''Regulatory Uncertainty and Market Liquidity: The 2008 Short Sale Ban's Impact on Equity Option Markets'', \textit{Journal of Finance}, Vol66


\bibitem{Biswas}
Biswas G., Nikolova S. and Stahel C., 2015, ''The Transaction Costs of Trading Corporate Credit'', \textit{SSRN}. 


\bibitem{Bjork}
Bjork, T., 2004, \textit{Arbitrage Theory in Continuous Time}, Oxford Univ. Press.


\bibitem{Brigo}
Brigo, D., Morini M. and Pallavicini A., 2013, \textit{Counterparty Credit Risk, Collateral and Funding: With Pricing Cases for All Asset Classes}, John Wiley and Sons Ltd.  

\bibitem{BrigoCapponi}
Brigo, D., Capponi A., Pallavicini A., 2014, ''Arbitrage-free Bilateral Counterparty risk valuation under collateralization and application to credit default swaps'', \textit{Mathematical Finance}, Vol:24.

\bibitem{BrigoIR}
Brigo, D., and Mercurio F., \textit{Interest Rate Models Theory and Practice, with Smile, Inflation and Credit}, 2nd Edition, Springer-Verlag, Page 775. 

\bibitem{BrigoNVA}
Brigo, D., Francischello, M., and Pallavicini, A., November 2015, ''Invariance, existence and uniqueness of solutions of nonlinear valuation PDEs and FBSDEs inclusive of credit risk, collateral and funding costs'', \url{https://arxiv.org/abs/1506.00686}.

\bibitem{2018a}
Brummelhuis, R. and Z. Luo. 2018a. ''CDS Proxy Construction via Machine Learning Techniques: Methodology and Results''. forthcoming, \textit{Journal of Financial Data Science}.   

\bibitem{2018b}
Brummelhuis, R. and Z. Luo. 2018b. ''CDS Proxy Construction via Machine Learning Techniques: Parametrization, Correlation and Benchmarking''. forthcoming, \textit{Journal of Financial Data Science}.   

\bibitem{2017}
Brummelhuis, R. and Z. Luo. 2017. ''CDS Rate Construction Methods by Machine Learning Techniques''. \textit{SSRN Journals.} \url{https://papers.ssrn.com/sol3/papers.cfm?abstract_id=2967184}. 

\bibitem{Burgard}
Burgard, C. and Kjaer, M., 2011, ''Partial differential equation representations of derivatives with bilateral counterparty risk and funding costs''. \textit{The Journal of Credit Risk} 7(3), 75–93.


\bibitem{Crepey}
Crepey, S., Gerboud, R., Grbac, Z., and Ngor, N., 2013, ''Counterparty risk and funding: The four wings of TVA''. \textit{Int'l Journal of Theoretical \& Applied Finance} 16, 1350006.

\bibitem{Delbaen}
Delbaen, F. and Schaechermayer, W., 1994, ''A General Version of the Fundamental Theorem on Asset Pricing''. \textit{Mathematishe Annalen} 300, 463-520

\bibitem{Schachermayer}
Delbaen, F. and W. Schachermayer, 2006, \textit{The Mathematics of Arbitrage. Springer Finance}, xvi+371 p., ISBN 3-540-21992-7 (2006).


\bibitem{GreenKenyonDennis}
Green, A., Kenyon C. and Dennis C., 2014, ''KVA: Capital Valuation Adjustment'', \textit{Risk}, 27.12.

\bibitem{GreenKenyon}
Green, A. and C. Kenyon. 2015. ''MVA: Initial Margin Valuation Adjustment by Replication and Regression''. \textit{Risk} 28(5).


\bibitem{Jarrow2}
Jarrow R., Li H, Ye X. and Hu M., 2018, ''Exploring Mispricing in the Term Structure of CDS Spreads'', forthcoming, Review of Finance, rfy014, https://doi.org/10.1093/rof/rfy014

\bibitem{Kapadia}
Kapadia N. and Pu X., 2012, ''Limited arbitrage between equity and credit markets'', \textit{Journal of Financial Economics} 105 (2012) 542564
 
 
\bibitem{Okane1}
O'Kane, D., 2008, \textit{Modelling single-name and multi-name Credit Derivatives}, John Wiley \& Sons Ltd.



 
\end{thebibliography}
\end{document}